\documentclass[aps,prb,amsfonts,amssymb,longbibliography,twocolumn,amsmath,floatfix,showpacs]{revtex4-1}
\usepackage{amsbsy}
\usepackage{subfig}
\usepackage{subfloat}
\usepackage[dvips]{color}
\usepackage[dvips]{graphics}
\usepackage{graphicx} 
\usepackage{bm} 
\usepackage[T2A]{fontenc}
\usepackage[cp1251]{inputenc}
\usepackage[english]{babel}

\newcommand{\be}{\begin{equation}}
\newcommand{\ee}{\end{equation}}
\newcommand{\bel}{\begin{align}}
\newcommand{\eel}{\end{align}}
\newcommand{\bem}{\begin{multline}}
\newcommand{\eem}{\end{multline}}
\newcommand{\beq}{\begin{equation}}
\newcommand{\eeq}{\end{equation}}
\newcommand{\bea}{\begin{eqnarray}}
\newcommand{\eea}{\end{eqnarray}}

\DeclareMathOperator{\sgn}{sgn}

\begin{document}

\title{Phase relations in superconductor-normal metal-superconductor tunnel junctions}

\author{Yu.\,S.~Barash}

\affiliation{Institute of Solid State Physics of the Russian Academy of Sciences,
Chernogolovka, Moscow District, 2 Academician Ossipyan Street, 142432 Russia}

\date{January 2, 2019}

\begin{abstract}
The phase difference $\phi$, between the superconducting terminals in 
superconductor-normal metal-superconductor tunnel junctions (SINIS), incorporates the phase 
differences~$\chi_{1,2}$~across~thin interfaces of constituent $SIN$ junctions and the phase incursion 
$\varphi$ between the side faces of the central electrode of length $L$. It is demonstrated here that 
$\chi_{1,2}$ pass through over their proximity-reduced domain twice, there and back, while $\phi$ changes
over the single period. Two corresponding solutions, that describe the double-valued order-parameter 
dependence on $\chi_{1,2}$, jointly form the single-valued dependence on $\phi$, operating in two adjoining 
regions of $\phi$. The phase incursion $\varphi$ plays a crucial role in creating such a behavior. The 
current-phase relation $j(\phi,L)$ is composed of the two solutions and, at a fixed small $L$, is 
characterized by the phase-dependent effective transmission coefficient.
\end{abstract}

\maketitle

{\it Introduction.}~When~two~superconductors~are~\mbox{separated} by a thin interface, their phase dependent
Josephson coupling generates the Josephson supercurrent through the junction. \cite{Josephson1962,%
Josephson1964,Josephson1965} If a normal metal is placed between the superconductors with a nonzero 
interface transparencies (SINIS), their Josephson coupling appears as a corollary of the proximity-induced 
superconducting correlations in the normal-metal region. \cite{deGennes1964,Likharev1979,Belzig1999,%
Klapwijk2004,Golubov2004} Various hybrid systems, in which the Josephson coupling through normal metal 
electrodes is induced by the proximity effect, have recently been the focus of research activities. 
\cite{Petrashov1995,Devoret1996,Schoen2001,Belzig2002,Esteve2008,Giazotto2010,Giazotto2011_2,Giazotto2011,%
Giazotto2014,Giazotto2015,Giazotto2017,Ryazanov2017} 

\begin{figure}[t]
\centering
\includegraphics*[width=.8\columnwidth,clip=true]{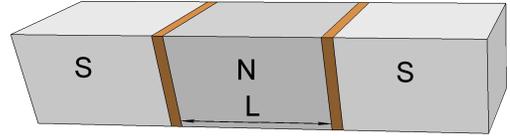}
\caption{Schematic diagram of the SINIS junction} \label{fig1}
\end{figure}

A SINIS junction (see Fig.\,\ref{fig1}) is characterized by the phase difference $\phi$ between the 
superconducting terminals, which can be represented as the sum of internal phase differences $\chi_{1,2}$, 
across the interfaces of two constituent $SIN$ junctions, and the supercurrent-induced phase incursion 
$\varphi$ between side faces of the central electrode of length $L$: $\phi=\chi_{1}+\chi_{2}+\varphi$.
This Rapid Communication develops a theory of symmetric SINIS tunnel junctions within the Ginzburg-Landau (GL) approach
that~allows~one~to~describe~the proximity effects of the Josephson origin on the phases mentioned 
above, and consequently on the junction's characteristics. 

For the junctions in question, one usually gets $\chi_{1,2}=\chi$ in equilibrium. The internal phase 
difference $\chi$ could~be controlled by the magnetic flux through an auxiliary superconducting ring 
involving only the constituent SIN contact, where the normal metal lead is in the proximity-induced 
superconducting state. However, it is $\phi$ that is commonly used as a control parameter in experiments 
and establishes both $\chi(\phi,L)$ and $\varphi(\phi,L)$. It will be demonstrated below that \!$\chi$, 
as opposed to $\phi$, does not determine the junction state uniquely at a given $L$. For this reason 
$\phi(\chi,L)$ and the order parameter absolute value~represent the double-valued functions of $\chi$.

There are two \mbox{solutions} to the GL equation that come up since the equation cannot be linearized, 
even when the order parameter is very small in the given problem. Such a linearization is known to represent 
the simplest and most effective way of describing the problems of $H_{c2}$ and $H_{c3}$ \cite{Abrikosov1957,%
DeGennes1963,Tinkham1996} as well as some proximity effects in the vicinity of superconductor-normal metal 
boundaries \cite{DeGennes1969,Abrikosov1988}. However, the linearization becomes impossible in the presence 
of a sizeable gauge invariant gradient of the order parameter phase, i.e., the superfluid velocity. After
switching over from $\chi$ to $\phi$, the two found solutions operate in different regions $|\phi|\le
\phi_*(L)$ and $\phi_*(L)\le|\phi|\le\pi$, within the period, adjoining at the points $\phi=\pm\phi_*(L)$. 
The phase incursion $\varphi$ plays a crucial role in creating such a behavior. As a corollary, the 
current-phase relation $j(\phi,L)$ is composed of the two solutions and its dependence on the transparency,
at small $L$, gradually changes with $\phi$ due to the phase incursion effects.

The influence of interfacial proximity effects in \mbox{SINIS} junctions on the phase relations, that results 
in a nonmonotonic dependence $\chi(\phi,L)$ at a fixed $L$,~has~not~been identified in the literature until 
now.
\begin{figure*}%
\centering
\subfloat[][]{\includegraphics[height=0.55\columnwidth]{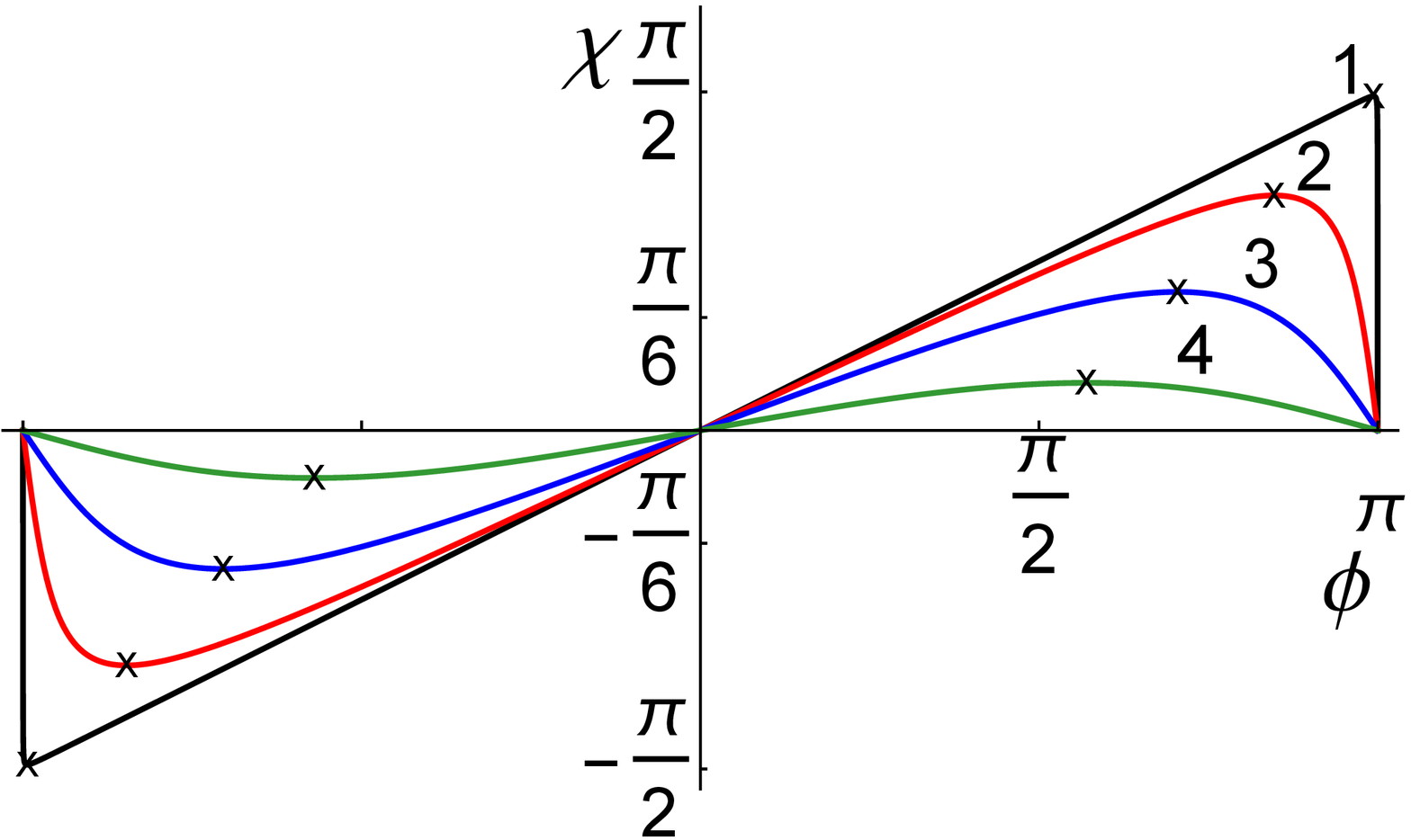}\label{fig:chi-phi-l}}%
\hspace{25mm}
\subfloat[][]{\includegraphics[height=0.55\columnwidth]{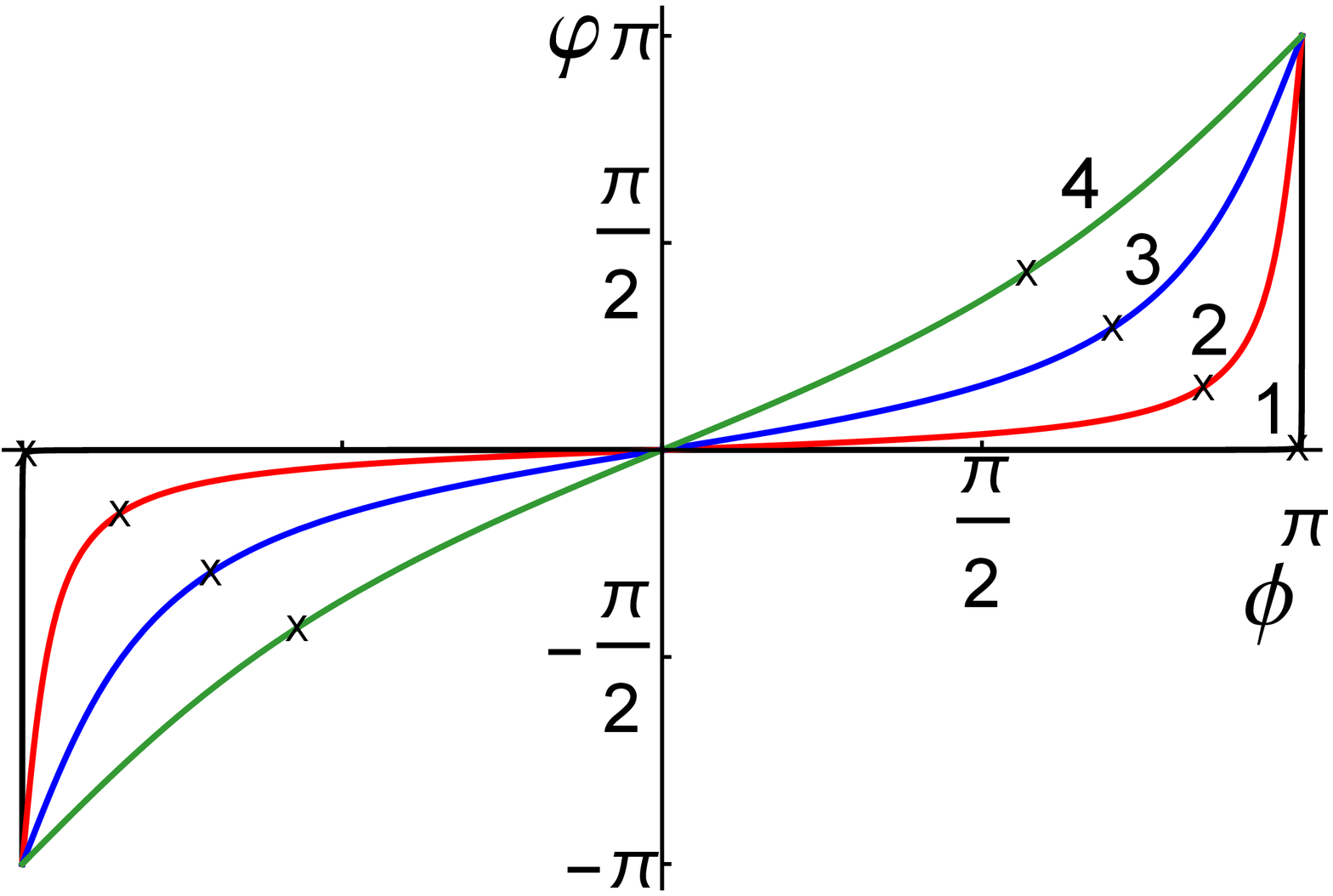}\label{fig:varphi-phi-l}}%
\caption{The internal phase difference $\chi$ (a) and the phase incursion $\varphi$ (b)
as functions of the phase difference $\phi$ taken at various distances $l$:\,
(1)\, $l=0.02$\,\, (2)\, $l=0.5$\,\, (3)\, $l=1.1$\,\, (4)\, $l=2.2$.}%
\label{fig:chi-varphi-phi-l}%
\end{figure*}
The \mbox{relation} was typically simplified assuming negligible values of either the phase incursion 
$\varphi$ over the central lead, or the phase drops $\chi$ across thin interfaces. More advanced earlier 
attempts of describing the SNS junction within the GL approach \cite{VolkovA1971,Fink1976} focused on the 
phase incursion and fully transparent interfaces, but were based on the specific boundary conditions and 
gave no consideration to the phase drops. \cite{Note1} On the other hand, 
microscopic theories of the double-junction systems, being elaborated on and applied to a wide temperature 
range down to zero temperature \cite{Kupriyanov1988,Kupriyanov1999,Golubov2000,Golubov2004}, usually assume a 
negligible current-induced phase incursion in contrast to the phase drops. While the latter point is 
justified for SISIS junctions in a wide range of realistic parameters, the range gets narrower in SINIS 
junctions, allowing both quantities $\chi$ and $\varphi$ to be of importance in the current transport 
at mesoscopic values of $L$, as shown below.

{\it Description of the model.}\,\,Consider a symmetric tunnel SINIS junction with two identical thin 
interfaces set at distance $L$ on the side faces of the central normal metal lead (see Fig.\,\ref{fig1}). 
A one-dimensional spatial profile of the order parameter will establish itself in the system, if, for example, 
the electrodes' transverse dimensions are much less than both the superconductor coherence length $\xi$ and 
the decay length $\xi_n$ in the central electrode. The system's free energy 
involves the bulk and interface contributions ${\cal F}=\sum{\cal F}_{p}+{\cal F}_{n}+
{\cal F}^{\text{int}}_{\frac{L}2}+{\cal F}^{\text{int}}_{-\frac{L}2}$. Here $p=1,2$ refer to the external 
superconducting electrodes, while subscript $n$ refers to the central normal metal lead. Assuming the latter 
to be described within~the GL approach \cite{Note2}, one gets per unit area of the cross section
\be
\!\!{\cal F}_n\!=\!\!\!\int\nolimits_{-L/2}^{L/2}\!\!\!dX\!
\left[\!K_n\left|\dfrac{d}{dX}\Psi(X)\right|^2\!\!\!+a_n\left|\Psi(X)\right|^2\!\!+\dfrac{b_n}{2}
\left|\Psi(X)\right|^4\right]\!,
\label{Fb1}
\ee
where $K_n,\,a_n,\,b_n>0$ and the interfaces are placed at $X=\pm L/2$. The expressions for ${\cal F}_{1,2}$ 
are obtained from \eqref{Fb1} after substituting $K_n,\,a_n,\,b_n\to K,\,-|a|,\,b$ and replacing the 
integration period $(-L/2,L/2)$ by $(-\infty,-L/2)$ or $(L/2,\infty)$ for $p=1$ or $2$, respectively.

The interfacial free energy per unit area is
\begin{equation}
{\cal F}^{\text{int}}_{\pm\frac{L}2}\!=g_{J}\left|\Psi_{\pm\frac{L}{2}+}\!-
\Psi_{\pm\frac{L}{2}-}\right|^2\!\!
+g\left|\Psi_{\pm\frac{L}{2}\pm}\right|^2\!\!+g_n\left|\Psi_{\pm\frac{L}{2}\mp}\right|^2\!.
\label{fint1}
\end{equation}
The first invariant in \eqref{fint1} describes the Josephson coupling while other terms take account of the 
interfacial pair breaking $g>0$, $g_n>0$.

The GL equation for the normalized absolute value of the order parameter in the central electrode
$\Psi=(a_n/b_n)^{1/2}fe^{\mathtt{i}\alpha}$ 
takes the form
\begin{equation}
\dfrac{d^2f}{dx^2}-\dfrac{i^2}{f^3}-f-f^3=0.
\label{gleq2}
\end{equation}
Here $x=X/\xi_n$, $\xi_n=(K_n/a_n)^{1/2}$ and the dimensionless current density is 
$i=\frac{2}{3\sqrt{3}}(j\big/j_{\text{dp}})$, where $j_{\text{dp}}=
\bigl(8|e|a_n^{3/2}K_n^{1/2}\bigr)\big/\bigl(3\sqrt{3}\hbar b_n\bigr)$. 

The boundary conditions for the complex order parameter, which follow from \eqref{Fb1} and \eqref{fint1}, 
agree with the microscopic results \cite{Golubov2004} near $T_c$, at all transparency values 
\cite{Barash2012,Barash2012_2,Barash2014_3}. Introducing $l=L/\xi_n$, one gets, in particular:
\begin{align}
&\left(\dfrac{df}{dx}\right)_{l/2-0}\!\!=
-\Bigl(g_{n,\delta}+g_\ell\Bigr)f_{-}+ g_\ell\cos\chi f_{+},
\label{bc1} \\
&i=-\,f^2\dfrac{d\alpha}{dx}=g_{\ell}f_{-}f_{+}\sin\chi.
\label{joscurr1}
\end{align}
Here $\chi\!=\!\alpha\left(\frac{l}{2}-0\right)\!-\!\alpha\left(\frac{l}{2}+0\right)$,\,
$f_{-}\!=\!f_{l/2-0}$,\, $f_{+}\!=\!f_{l/2+0}$
and the dimensionless coupling~constants~$g_\ell\!=\!g_J\xi_n/K_n$, $g_{n,\delta}=g_n\xi_n/K_n$.

{\it The phase relations.} The proximity effect of the \mbox{Josephson} origin, associated with the term 
containing $\cos\chi$ in~\eqref{bc1}, takes place under the condition $g_\ell\cos\chi>0$, when the
Josephson-coupling bilinear contribution to the free energy $\propto-2g_\ell f_-f_+\cos\chi$ 
decreases with an appearance of a small nonzero order parameter $f_-$ on one side~of a thin 
interface, in the presence of $f_+$ on the other side. For $0$-junctions considered below $g_\ell>0$.
Therefore, for superconductivity to appear in the central lead, the internal phase difference $\chi$ should 
change within the proximity-reduced range, $|\chi|\le\chi_{\text{max}}(l)<\frac{\pi}{2}$, which is defined, 
in general, modulo $2\pi$. If $\chi$ were outside the range, the inverse proximity effect would prevent
superconductivity to show up in the central lead.

\begin{figure*}%
\centering
\subfloat[][]{\includegraphics[height=0.55\columnwidth]{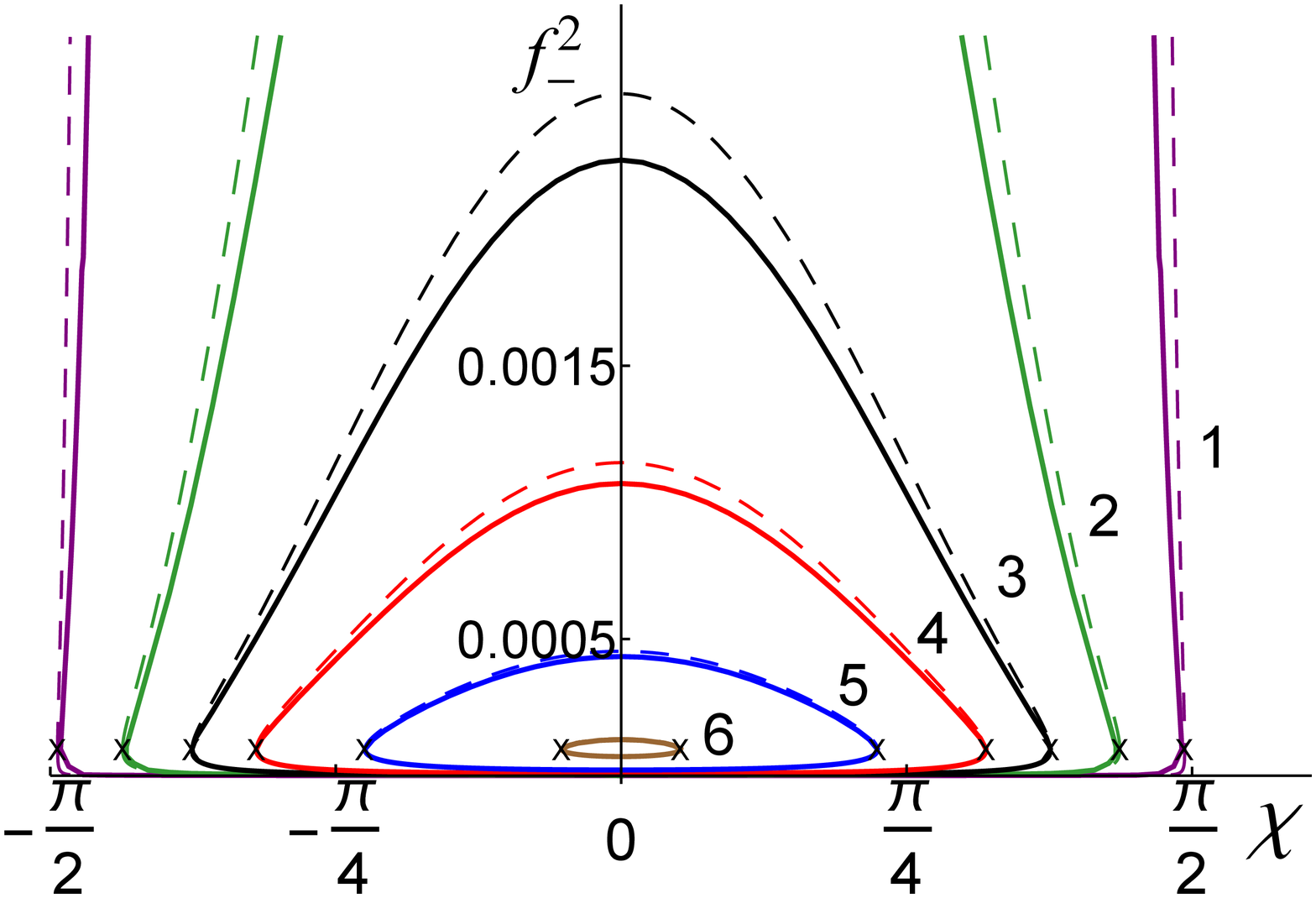}\label{fig:t0-chi-l}}%
\hspace{25mm}
\subfloat[][]{\includegraphics[height=0.55\columnwidth]{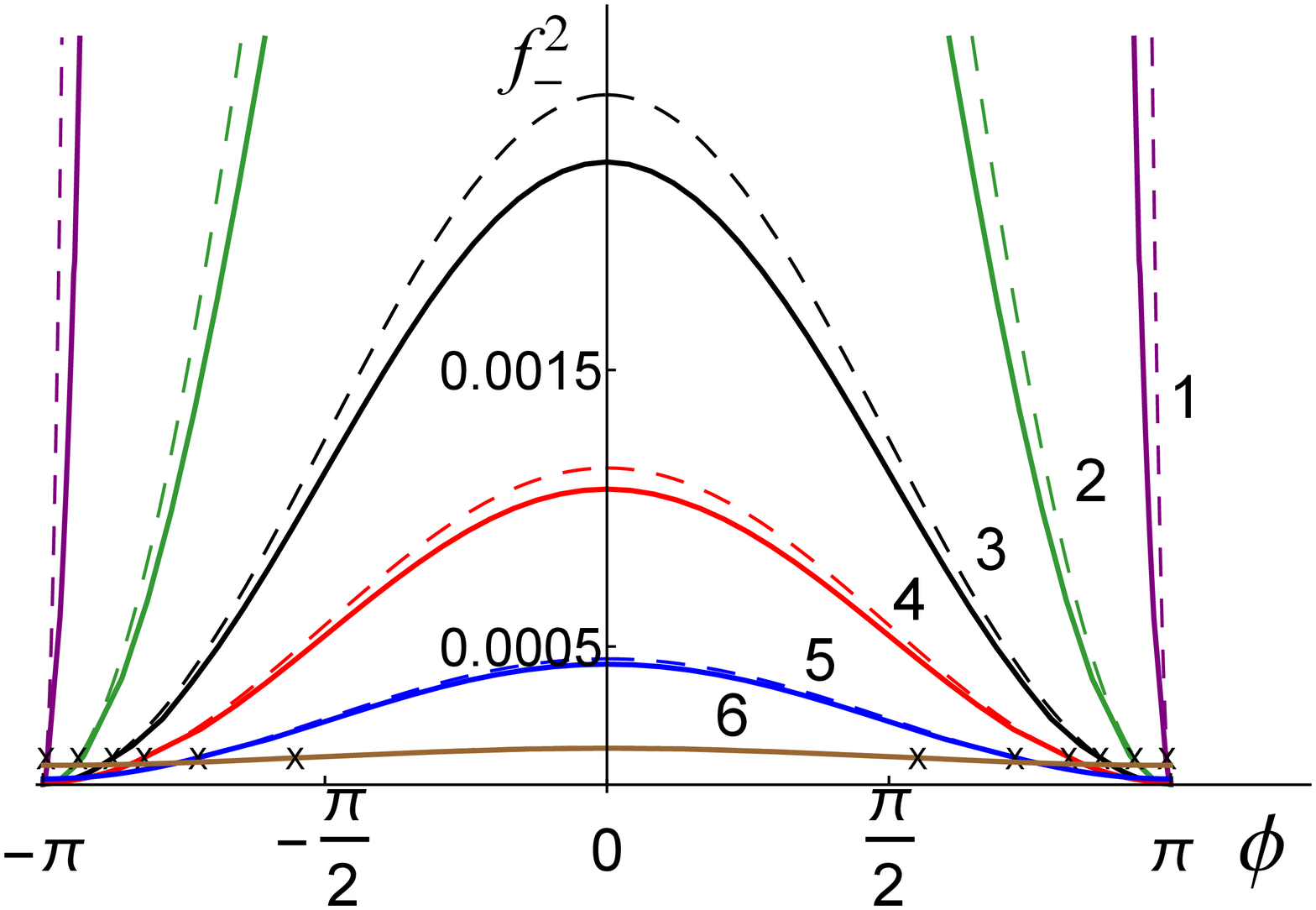}\label{fig:t0-phi-l}}%
\caption{The quantity $f_{-}^2$ as (a) a double-valued function of $\chi$  and (b) a single-valued 
function of $\phi$ taken at various $l$:\, 
(1)\, $l=0.02$\,\, (2)\, $l=0.2$\,\, (3)\, $l=0.4$\,\, (4)\, $l=0.6$\,\, (5)\, $l=1$\,\, 
(6)\, $l=2.5$.}%
\label{fig:t0-chi-phi-l}%
\end{figure*}

Figs.\,\ref{fig:chi-phi-l} and \ref{fig:varphi-phi-l} show, respectively, the internal phase difference $\chi$
and the phase incursion $\varphi$, taken at various $l$ as functions of the phase difference $\phi$ between 
the superconductor terminals. All the numerical results~have~been obtained by carrying out an evaluation of 
the GL equations' solutions that take into account the phase incursion and boundary conditions at interfaces 
with $g_\ell\!=\!g_\delta\!=\!0.01$ and $g_{n,\delta}\!=\!0$, assuming $K=K_n$, 
$|a|=a_n$ and $b=b_n$. \!The approximate analytical \mbox{solutions}~for~tunnel SINIS junctions have been also
obtained. \cite{Note3} \!They describe 
almost perfectly the functions $\chi(\phi)$ and $\varphi(\phi)$ for the parameter set chosen, with deviations 
from the numerical results that are indiscernible in Figs.\,\ref{fig:chi-phi-l} and \ref{fig:varphi-phi-l}.

If the phase incursion $\varphi$ is negligibly small, one gets from $\phi=2\chi+\varphi$ a simple dependence 
$\chi(\phi)=\frac{\phi}2$, which results in the variation range $|\chi|\le\frac{\pi}{2}$ for $|\phi|\le\pi$. 
Such a behavior takes place at sufficiently small distances $l\ll1$, except for a narrow vicinity of $\phi=
\pi$, as shown in the curves 1 in Figs.\,\ref{fig:chi-phi-l} and \ref{fig:varphi-phi-l}. The curves 2-4 
demonstrate that, in a wide range of $\phi$, $\chi$ is of importance at mesoscopic lengths $l\alt1$, while a 
substantial influence of $\varphi$ on the phase relations appears at $l\agt1$.

Since the supercurrent is spatially constant due to the presumed quasi-one-dimensional character of the 
problem, a local decrease of the Cooper pair density is accompanied by the increase of the 
superfluid velocity, i.e., of the gradient of the order parameter phase. Therefore, small local values of 
$f$ result in a noticeable $\varphi$. Due to a spatial decay of the proximity-induced condensate density 
with increasing distances from the interfaces, $\varphi$ increases with $l$ at a given $\phi$, 
while the range of variation of $|\chi|\le\chi_{\text{max}}(l)$ becomes smaller:
$\chi_{\text{max}}(l)\approx\arccos(\tanh l)$. At $l\gg1$ $f$ is especially small in the depth of the central 
electrode, $\varphi$ dominates the right-hand side in $\phi=2\chi+\varphi$, while $|\chi|$ is greatly reduced. 

Fig.\,\ref{fig:chi-phi-l} demonstrates that $\chi$ is a nonmonotonic function of $\phi$ that passes through
over the proximity-reduced region twice, there and back, while the phase difference $\phi$ between the 
superconducting terminals changes over the period. Two different values of $\phi$ at one and the same $\chi$ 
are linked to the different phase incursions and, more generally, to the two solutions of the GL equation for 
the absolute value of the order parameter, taken at a given $\chi$. The dots marked with crosses represent in 
all the figures the points of contact of the two solutions, i.e., indicate the corresponding quantities taken 
at $\chi=\pm\chi_{\text{max}}(l)$. 

\begin{figure*}%
\centering
\subfloat[][]{\includegraphics[height=0.57\columnwidth]{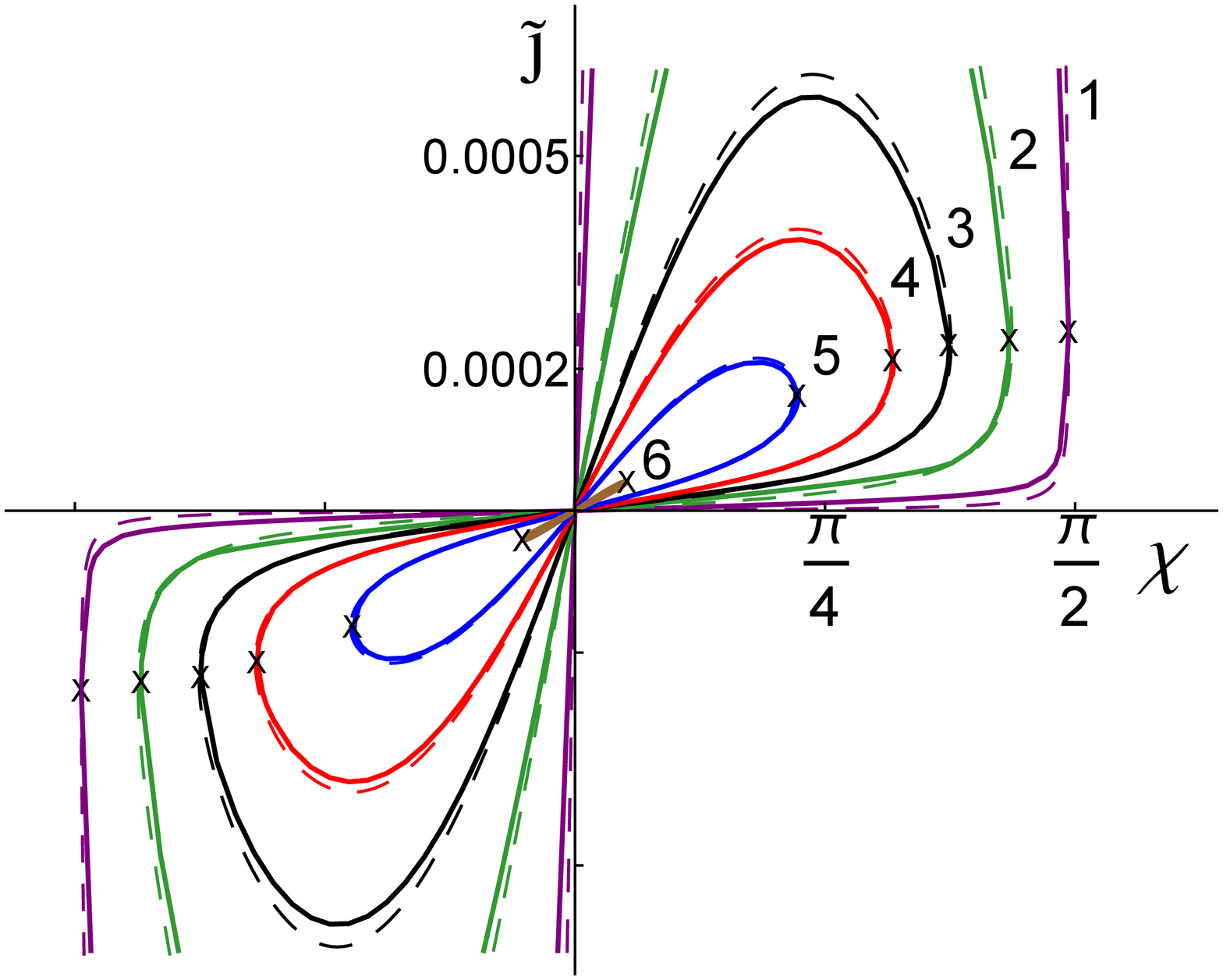}\label{fig:cur-chi-l}}%
\hspace{20mm}
\subfloat[][]{\includegraphics[height=0.57\columnwidth]{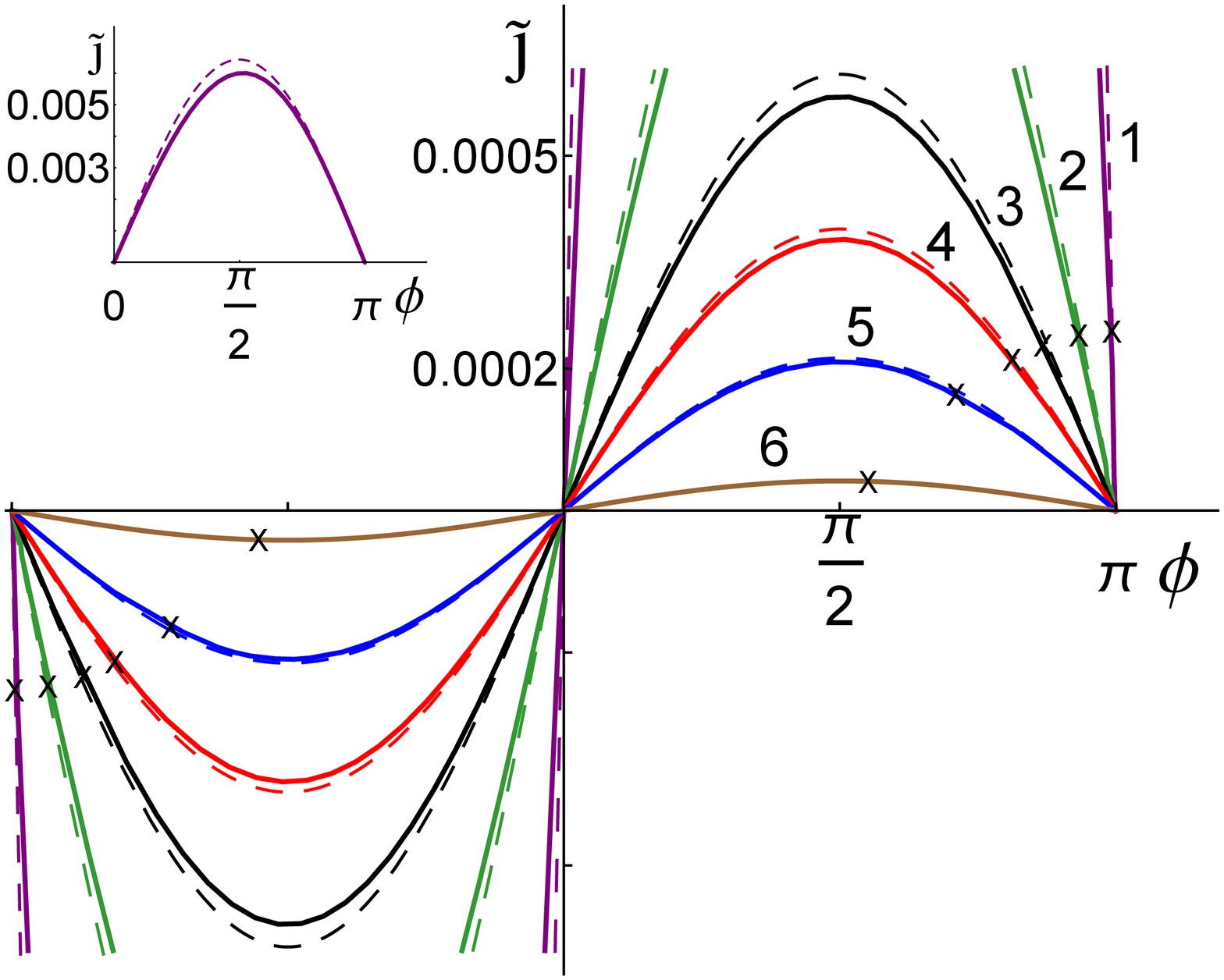}\label{fig:cur-phi-l}}%
\caption{Normalized supercurrent as (a) a double-valued function of $\chi$ and (b) a single-valued 
function of $\phi$ taken at various $l$: \, (1)\, $l=0.02$\,\, (2)\, $l=0.2$\,\, 
(3)\, $l=0.4$\,\, (4)\, $l=0.6$\,\, (5)\, $l=1$\,\, (6)\, $l=2.5$.\\ 
Inset: The supercurrent at l=0.02 (solid line) and its analytical description at small $l$ (dashed line).}%
\label{fig:cur-chi-phi-l}%
\end{figure*}

{\it The order parameter $f_-$.}\,\,
The nonlinear term $i^2f^{-3}\propto v_s^2(x)f(x)$ (where the superfluid velocity is $v_s(x)\propto i/f^2$) 
cannot, as a rule, be disregarded in \eqref{gleq2} as compared to the linear one. In the depth of the central
lead it dominates the latter, when $\phi$ is close to $\pi$. For this reason the GL equation \eqref{gleq2}
remains nonlinear even if the cubic term is negligible in the problem under consideration. As a result, there 
are two basic solutions for $f$ at a given $\chi$.

The normalized order-parameter absolute value squared $f^2_-$ taken at a side face of the central electrode 
is shown in Figs.\,\ref{fig:t0-chi-l} and \ref{fig:t0-phi-l} at various $l$ as a function of $\chi$ and 
$\phi$, respectively. The analytical description (dashed curves), that assumes the conditions $f_-\!\sim 
\!g_\ell f_+\!\ll\!1$~and $g_{n,\delta}\alt g_\ell$, approximates the numerical results shown reasonably well.
As distinct from the phase relations~in Fig.\,\ref{fig:chi-varphi-phi-l}, the solid and dashed curves in 
Fig.\,\ref{fig:t0-chi-phi-l} can be, mostly, clearly distinguished. The two solutions adjoin~at $\chi\!=\!\pm
\chi_{\text{max}}(l)$ and form the double-valued behavior shown in Fig.\,\ref{fig:t0-chi-l}. The 
first solution for $f_-$ has a maximum and the second one a minimum at $\chi\!=\!0$ at a fixed $l$. The~same 
occurs at $l\to0$ at a fixed $\chi$, where the minimum is zero. 

If $\chi$ were fixed experimentally, the first solution would describe the stable and the second one the 
metastable states. However, the control parameter in experiments is usually $\phi$. After switching over from 
$\chi$ to $\phi$, the order parameter is described by the continuous single-valued dependence $f_-(\phi,l)$ 
shown in Fig.\,\ref{fig:t0-phi-l}. The first solution operates in the region $|\phi|\in(0,\phi_*(l))$ while 
the other one is in $|\phi|\in(\phi_*(l),\pi)$. Here $\phi_*(l)\approx\frac{\pi}{2}+\arcsin(\frac{1}{\cosh l})$.
The adjoining regions do not overlap due to a substantial phase incursion occurring at small $f$. The curves'
crossing, seen in Fig.\,\ref{fig:t0-phi-l} at small $f_-$, is a manifestation of the opposite behavior of the 
two solutions with increasing $l$. If $\phi=\pi$, $f$ is zero at $x=0$ at arbitrary $l$, that allows 
phase-slip processes in the central lead. \cite{Fink1976,Giazotto2017,Note3}

For tunnel interfaces one obtains $f_-\sim g_\ell f_+\ll1$ at $g_{n,\delta}\alt g_\ell$,
except for the first solution at sufficiently small $l$. The latter results, in the limit $l\to0$,~in the 
relation $f_-\!=\!\frac{g_\ell\cos\chi}{g_\ell+g_{n,\delta}}f_+$, which also applies to SISIS junctions 
\cite{Barash2018} and approximately describes the dependence on $\chi$. For the whole parameter set used in 
the figures $f_+$ weakly changes with $\chi$ and $l$:~$f_+^2\in(0.972,0.978)$. If $g_{n,\delta}\alt g_\ell$ 
and $\cos\chi\sim1$, one obtains $f_-\sim f_+$, while in the opposite case $g_{n,\delta}\gg g_\ell$ the 
relation is $f_-\ll f_+$. Since $g_\ell$ for tunnel interfaces is proportional to the 
transmission coefficient $g_\ell\propto{\cal D}$ \cite{Barash2012,Barash2012_2,Barash2014_3}, the above 
relation results in $f_-\propto{\cal D}$,~if $g_\ell\ll g_{n,\delta}$, and in the ${\cal D}$-independent 
quantity $f_-$~for $g_\ell\gg g_{n,\delta}$. The second solution vanishes in the limit $l\to0$, and satisfies 
the relation $f_-=g_\ell f_+\tanh\frac{l}{2}$ at arbitrary $l$ and $\chi=0$. In the case of large $l$ the two 
solutions coincide and the relation at $\chi=0$ is $f_-=g_\ell f_+$. \cite{Note3} 

{\it The supercurrent.}\,
The normalized supercurrent $\tilde{\jmath}=j\big/j_{\text{dp}}$ is depicted in Figs.\,\ref{fig:cur-chi-l} and
\ref{fig:cur-phi-l} at various $l$ as a function of $\chi$ and $\phi$, respectively.\!\! With respect to
$\chi$, the supercurrent is in the shape of a double loop that looks like a sloping figure eight composed of
the two solutions. After switching over from $\chi$ to $\phi$ the current-phase relation acquires the 
conventional form. The dashed curves, that correspond to approximate analytical results \cite{Note3}, have the
sinusoidal shape in Fig.\,\ref{fig:cur-phi-l}. They deviate within several percent from the numerical results 
(solid curves).

A substantial role of the phase incursion in creating such a behavior can be understood as follows. The 
supercurrent $\propto g_{\ell}f_{-}f_{+}\sin\chi$ is influenced by the proximity effect together with
$f_-$. If $\varphi$ were completely neglected, the value
$\phi\!=\!\pi$ would correspond to $\chi\!=\!\frac{\pi}2$. Since the proximity effect vanishes
at $\chi\to\frac{\pi}{2}$, one gets $f_-\!\!\to0$ that could explain the zeroth supercurrent at $\phi=\pi$. 
However, as $f_-$ is small in the vicinity of 
$\phi=\pi$, one gets a noticeable phase incursion that reduces the variation range $|\chi|\le
\chi_{\text{max}}(l)<\frac{\pi}{2}$ and excludes a possibility for $\chi$ to reach $\frac{\pi}{2}$ at any 
nonzero $l$. Instead, there appear two solutions of the GL equation providing a return passage for $\chi$, 
from $0$ to $\chi_{\text{max}}(l)$ and back, while $\phi$ changes over $(0,\pi)$. As a result, the 
correspondence of $\chi=0$ to both $\phi=0$ and $\phi=\pi$ is established. The phase relations in the 
SINIS systems do not result in the regime of interchanging modes with abrupt supercurrent changes, that 
can occur in SISIS junctions. \cite{Luca2009,Linder2017,Barash2018}

The small values of the order parameter and supercurrent, that are characteristic for the second solution and 
marked with crosses in Figs.\,\ref{fig:t0-chi-phi-l} and \ref{fig:cur-chi-phi-l}, are specifically associated 
with the choice~$g_\ell=0.01$ for the Josephson coupling constant, taken for demonstrating a quantitative 
agreement between the numerical and approximate analytical results. The effects discussed increase with 
$g_\ell$ and remain qualitatively the same for $g_\ell\alt 1$. \cite{Note4} Thus for 
$g_\ell=0.1$ instead of $g_\ell=0.01$, the characteristic values of $f_-^2$ and $\tilde{\jmath}$
increase in about $50-100$ times.

The first solution is strongly modified at small distances $l\alt 2g_\ell (b_n|a|/ba_n)^{1/2}\ll1$,
for which the analytical description, based on the relation $f_-=\frac{g_\ell\cos\chi}{g_\ell+
g_{n,\delta}}f_+$ rather than on $f_-\sim g_\ell f_+\ll1$, has to be developed. \cite{Note3} The inset in
Fig.\,\ref{fig:cur-phi-l} shows 
the solid curve 1 as a whole ($l=0.02$). The analytical results deviate
weakly from the solid curve. When $g_{n,\delta}\ll g_\ell$, one obtains $j\propto {\cal D}$.
A remarkable feature is that the supercurrent dependence on the 
transparency gradually evolves into ${\cal D}^2$ with increasing $\phi$ up to about $\phi_*$ at a fixed
small $l$, due to the increase of the phase incursion with $\phi$. Similar supercurrent behavior also 
takes place for some other reasons, in particular, with the increasing distance $l$. \cite{Kupriyanov1988,%
Kupriyanov1999,Golubov2000,Golubov2004} For the second solution one always obtains $j\propto{\cal D}^2$. 
Such a crossover is a fingerprint of the underlying physics associated with the phase-dependent proximity 
effect of the Josephson origin that generates the unconventional behavior of internal phase differences in 
SINIS junctions.

The research is carried out within the state task of ISSP RAS.



\onecolumngrid
\clearpage
\begin{center}
{
{\large \textbf{Phase relations in superconductor-normal metal-superconductor tunnel junctions}}
\\[10pt]
\textbf{Supplemental Material}
}
\thispagestyle{empty}

\vspace{0.15in}

Yu.\,S.~Barash

\vspace{0.075in}
\small\textit{Institute of Solid State Physics of the Russian Academy of Sciences,
Chernogolovka, Moscow District, 2 Academician Ossipyan Street, 142432 Russia}

\end{center}

\vspace{3ex}

\numberwithin{equation}{section}

\vspace{-0.1cm}
Two symmetric one-dimensional solutions to the GL equation are analytically described for tunnel SINIS 
junctions with the parameters linked by the boundary and asymptotic conditions. 
\vspace{0.7cm}

\twocolumngrid
\setcounter{subsection}{0}
\setcounter{equation}{0}
\renewcommand{\thesubsection}{S\arabic{subsection}}
\renewcommand{\theequation}{S\arabic{equation}}

\subsection{Symmetric solutions to the GL equation}
\label{sec: sol}

The GL equation for the absolute order-parameter value can be written as 
\be\left\{
\begin{aligned}
&\dfrac{d^2f}{dx^2}-\dfrac{i^2}{f^3}-f-f^3=0,\quad |x|<l/2,\\  \\
&\dfrac{d^2f}{dx^2}-\dfrac{K_n^2}{K^2}\cdot\dfrac{i^2}{f^3}+\dfrac{|a|K_n}{a_nK}f-
\dfrac{b K_n}{b_nK}f^3=0,\quad |x|>l/2.
\end{aligned}
\right.
\label{rlambda11}
\ee
Here the dimensionless quantities $f$ and $x$ are defined as $\Psi\!=\!(a_n/b_n)^{1/2}fe^{\mathtt{i}\alpha}$,
$x\!=\!X/\xi_n$. Also $l\!=\!L/\xi_n$, $\xi_n\!\!=\!(K_n/a_n)^{1/2}$ and the dimensionless 
current density is $i\!=\!\frac{2}{3\sqrt{3}}(j\big/j_{\text{dp}})$, where $j_{\text{dp}}\!=\!
\bigl(8|e|a_n^{3/2}K_n^{1/2}\bigr)\big/\bigl(3\sqrt{3}\hbar b_n\bigr)$. One also assumes $a_n\sim|a|$ that
makes possible a joint description of the normal metal and superconducting leads within the GL approach. 
\cite{SAbrikosov1988}

The boundary conditions at the interfaces at $x=\pm l/2$ take the form
\be\left\{
\begin{aligned}
&\left(\dfrac{df}{dx}\right)_{\pm(l/2-0)}=\mp\bigl(g_{n,\delta}+g_{\ell}\bigr)f_-\pm g_{\ell}f_+\cos\chi,
\\
&\dfrac{K}{K_n}\left(\dfrac{df}{dx}\right)_{\pm(l/2+0)}=\pm(g_\delta+g_{\ell})f_+\mp g_{\ell}f_-\cos\chi,
\end{aligned}
\right.
\label{rlambda30a}
\ee
where the dimensionless coupling constants are $g_\ell\!=\!g_J\xi_n(T)/K_n$, $g_{n,\delta}=g_n\xi_n(T)/K_n$
and the symmetric solutions $f(x)=f(-x)$ are considered.

The Josephson current described by the relation (5) of the main text
can also be expressed via the asymptotic absolute value $f_\infty$ of the superconductor order parameter:
\be
i^2=\dfrac{Kb}{K_nb_n}\left(\dfrac{|a|b_n}{a_nb}-f_\infty^2\right)f_\infty^4.
\label{ascur1}
\ee
In the absence of the supercurrent one gets, for the normalization chosen, 
\be
f_{\infty}^2=\frac{|a|b_n}{a_nb}.
\label{as1}
\ee

Equating (5) and \eqref{ascur1} results in the equation
\be
\dfrac{Kb}{K_nb_n}\left(\dfrac{|a|b_n}{a_nb}-f_\infty^2\right)f_\infty^4=
g_\ell^2f_-^2f_+^2\sin^2\chi.
\label{sol3}
\ee

Symmetric analytical solutions to the GL equations \eqref{rlambda11} describe the order-parameter absolute
value that satisfies the boundary conditions \eqref{rlambda30a} at the thin 
interfaces, as well as the asymptotic conditions \eqref{as1} deep inside the long superconductor leads. The 
energetically most favorable solutions are expected to describe the proximity-induced order parameter absolute
value in the normal metal lead with a single minimum at the center $x=0$ between the interfaces and the equal 
maximums $f_-$ at the side faces of the central electrode $x=\pm(l/2-0)$. In each of the external 
leads, the order parameter has its maximum $f_{\infty}$ at asymptotically large distances and minimum $f_+$
at pair breaking boundaries $x=\pm(l/2+0)$. The parameters $f_\pm$ and $f_\infty$ depend on the phase 
difference $\chi$ and the central lead's length $l$ and should be determined together with other parameters 
of the whole solution.

For obtaining the solutions, the first integral of \eqref{rlambda11} will be used. 
The corresponding quantities ${\cal E}_n$ and ${\cal E}$, defined as
\begin{align}
&{\cal E}_n=\!\left(\dfrac{df(x)}{dx}\right)^2\!\!\!+\dfrac{i^2}{f^2(x)}-f^2(x)-\nonumber\\
&\qquad \qquad \qquad \qquad \qquad -\dfrac{1}{2}f^4(x), \enspace |x|<l/2, \label{RPhi88gta}\\  
&{\cal E}=\left(\dfrac{df(x)}{dx}\right)^2+\dfrac{K_n^2}{K^2}\dfrac{i^2}{f^2(x)}+
\dfrac{|a|K_n}{a_nK}f^2(x)-\nonumber\\
&\qquad \qquad \qquad \qquad -\dfrac{bK_n}{2b_nK}f^4(x),\enspace |x|>l/2,
\label{RPhi88gt}
\end{align}
are spatially constant, when taken for the solutions to \eqref{rlambda11} inside the central electrode and
the external leads, respectively. Since the boundary conditions \eqref{rlambda30a} do not generally support
the conservation of ${\cal E}$ through the interfaces, ${\cal E}_n$ and ${\cal E}$ can substantially differ
from each other.

Eq. \eqref{RPhi88gta} can be rewritten in the form
\be
\left(\dfrac{dt}{dx}\right)^2=2(t-t_1)(t-t_2)(t-t_3),\enspace |x|<l/2,
\label{RPhi89gt}
\ee
where $t(x)\equiv f^2(x)$.

The quantities $t_i=f_i^2$, $i=1,2,3$ satisfy the following set of equations
\be
t_1+t_2+t_3=-2, \enspace 
t_1t_2t_3=2i^2, \enspace 
t_1t_2+t_1t_3+t_2t_3=2{\cal E}_n.
\label{ttt1}
\ee

Solutions to equation \eqref{RPhi89gt} are characterized by three formal extrema $t_1,\,t_2,\,t_3$ 
with the vanishing first derivative $\frac{dt}{dx}$. In general, either all three roots $t_1,\, t_2$ and 
$t_3$ take on real values, or only one is real and two are the complex conjugate of each other. As the 
numerical study shows, only real values of $t_i$ are relevant for the given problem. For superconducting leads 
real roots usually take nonnegative values $t_i\ge0$ ($i=1,2,3$). \cite{SBarash2017,SBarash2018}
For the proximity influenced normal metal lead only one of the real roots is nonnegative:\, $t_1\ge0$,\, 
$t_{2,3}\le0$, as ensured by the sign minus and sign plus on the right hand sides of the first and second 
equations in \eqref{ttt1}, respectively. The sign minus originates from the condition $a_n>0$ for the normal 
metal. For the superconducting lead $a<0$, that would result in the sign plus instead of minus. 

As the left 
hand side of \eqref{RPhi89gt} takes on nonnegative values, the quantity $t_1$ has to be the minimum. While the
negative roots $t_{2,3}$ correspond to purely imaginary quantities $f_{2,3}$, $t_1$ just represents the actual
minimum that $t(x)=f^2(x)$ takes at the central point of the normal metal lead $x=0$.

Thus, the solution inside the central lead $|x|<l/2$ should monotonically increase with $|x|$
and, in particular, have a nonnegative derivative at $x=l/2-0$. As follows from \eqref{rlambda30a}, 
the latter conditions require $-\bigl(g_{n,\delta}+g_\ell\bigr)\sqrt{t_-}+g_\ell\cos\chi \sqrt{t_+}\ge 0$ and
$t_3\le t_2\le0\le t_1\le t(x)\le t_-\le t_+$.
In accordance with the above, one gets from Eq. \eqref{RPhi89gt}:
\be
|x(t)|=\sqrt{\dfrac{2}{t_1-t_3}}F\left(\left.
\arcsin\sqrt{\frac{t-t_1}{t-t_2}}\,\right|\,\frac{t_2-t_3}{t_1-t_3}\right).
\label{sol1} 
\ee
Here the definitions of the Mathematica book are used for the notations of arguments of the elliptic integral
of the first kind $F\left(\varphi\left|\,m\right.\right)$. \cite{SWolfram2003}

Taking $x=l/2-0$ in \eqref{sol1} results in the condition associated with the central lead's length:
\be
\sqrt{\dfrac{2}{t_1-t_3}}F\!\left(\left.
\arcsin\sqrt{\frac{t_--t_1}{t_--t_2}}\,\right|\,\frac{t_2-t_3}{t_1-t_3}\right)=\dfrac{l}{2}.
\label{sol4}
\ee

The quantity ${\cal E}_n$ for the central lead can be expressed via $t_{\pm}$, taking $x=(l/2)-0$ 
in \eqref{RPhi88gta} and making use of (5) and the first equation in \eqref{rlambda30a}:
\begin{multline}
{\cal E}_n=\biggl[-1+\Bigl(g_{n,\delta}+g_\ell\Bigr)^2\biggr]t_-+g_\ell^2 t_+-\\
-2g_\ell\Bigl(g_\ell+g_{n,\delta}\Bigr)\cos\chi \sqrt{t_-t_+}-\dfrac{1}{2}t_-^2\,.
\label{boundcond54}
\end{multline}

Taking $x\to\infty$ in \eqref{RPhi88gt} and using \eqref{ascur1}, one obtains
\be
{\cal E}=\dfrac{bK_n}{b_nK}\left(2\dfrac{|a|b_n}{a_nb}-\dfrac32t_\infty\right)t_\infty.
\label{pap1}
\ee
On account of \eqref{ascur1} and \eqref{pap1}, equation \eqref{RPhi88gt} can be rewritten in the form
\be
\left(\dfrac{dt}{dx}\right)^2=\dfrac{2K_nb}{Kb_n}\bigl(t_\infty-t\bigr)^2
\left(t-2\left(\dfrac{|a|b_n}{a_nb}-t_\infty\right)\right).
\label{pap2}
\ee
The nonnegative value of the left hand side in \eqref{pap2} requires the solution $t(x)$ to satisfy the 
condition $t\ge2\left(\dfrac{|a|b_n}{a_nb}-t_\infty\right)$  in the external region $|x|>l/2$.

Taking the square root of both sides of \eqref{pap2} at $x=l/2+0$ and substituting the result in the 
second boundary condition in \eqref{rlambda30a}, one excludes the derivative of the order 
parameter absolute value, taken at the boundary, and obtains
\begin{multline}
\left(t_{\infty}-t_{+}\right)\sqrt{2t_{\infty}+t_{+}-2\frac{|a|b_n}{a_nb}}= \\ =
\sqrt{\dfrac{2K_nb_n}{Kb}}\left[\Bigl(g_\delta+g_\ell\Bigr)t_{+}-g_\ell\cos\chi \sqrt{t_{-}t_{+}}\right].
\label{calem10p}
\end{multline}
Positive sign of the left hand side in \eqref{calem10p} entails the relation
$\Bigl(g_\delta+g_\ell\Bigr)\sqrt{t_{+}}>g_\ell\cos\chi \sqrt{t_{-}}$. 

Since the supercurrent is expressed in (5) via the quantities $t_{\pm}$, the analysis of the full spatial 
order parameter profile can be omitted here. However, one needs to study the external 
phase difference $\phi=2\chi+\varphi$ and the associated phase incursion $\varphi$.

For the gradient of the order parameter phase one gets from (5) 
\be
\dfrac{d\alpha}{dx}=-\,\dfrac{i}{f^2}.
\label{pap3}
\ee
Integrating both sides in \eqref{pap3} along the central lead and using \eqref{RPhi89gt}, one has
\be
\varphi=\sqrt{2}i\!\!\int\limits_{t_1}^{t_-}\!\!\dfrac{dt}{t\sqrt{(t-t_1)(t-t_2)(t-t_3)}},
\label{pap4}
\ee
where $\varphi=\alpha(-l/2+0)-\alpha(l/2-0)$.
After taking into account the relation \eqref{sol4}, the result of the integration in \eqref{pap4} contains 
the elliptic integral of the third kind and takes the form
\begin{widetext}
\be
\varphi=\frac{2i}{t_2}\Biggl[-\,\dfrac{\sqrt{2}(t_1-t_2)}{t_1\sqrt{t_1-t_3}}\,
\Pi\!\left(\frac{t_2}{t_1-t_2};\arcsin\sqrt{\frac{t_--t_1}{t_--t_2}}
\left|\frac{t_2-t_3}{t_1-t_3}\right.\right)+\frac{l}{2}\Biggr].
\label{pap5}
\ee
\end{widetext}

The solutions to Eqs. \eqref{rlambda11} have to satisfy the relation 
\eqref{sol4}, the boundary and asymptotic conditions. As a result, the full parameter set contains six 
quantities $t_1$,\,$t_2$,\,$t_3$,\,$t_\pm$, and $t_\infty$, which  are linked to each other by six equations
\eqref{ttt1}, \eqref{sol4}, \eqref{calem10p} and \eqref{sol3}, where expressions \eqref{boundcond54} and 
(5) should be substituted for ${\cal E}_n$ and $i$. Joint solutions to the equations studied 
represent the parameters as functions of the phase difference $\chi$ and the dimensionless length $l$ of the 
central lead. Though the numerical study of such solutions is generally required, a number of important 
problems allow analytical descriptions. Both approaches demonstrate the presence of two solutions that satisfy
the above equations. The analytical solutions will be presented in the following sections.

\subsection{Exact results at $\chi=0$}
\label{sec: exact}

The description of currentless states in the SINIS junctions can be analytically reduced, for each of the
solutions, to a comparatively simple exact relation between $l$ and $t_-$. As it follows from 
(5) and \eqref{ascur1}, for the currentless states  $\chi=0$ and $t_\infty=\frac{|a|b_n}{a_nb}$.
With this $\chi$ and $t_\infty$, one gets from \eqref{calem10p} the expression for $f_+\equiv\sqrt{t_+}$ as 
a function of $f_-\equiv\sqrt{t_-}$:
\begin{multline}
f_+(f_-)=\sqrt{\dfrac{|a|b_n}{a_nb}+\dfrac{K_nb_n}{2Kb}(g_\delta+g_\ell)^2+\sqrt{\dfrac{2K_nb_n}{Kb}}g_\ell 
f_-}- \\ -\sqrt{\dfrac{K_nb_n}{2Kb}}(g_\delta+g_\ell).
\label{pap23}
\end{multline}

Taking $\chi=0$ in 
\eqref{ttt1}, \eqref{sol4}, where $i=0$ and ${\cal E}_n$ is 
defined in \eqref{boundcond54}, one obtains two solutions. The first solution corresponds to ${\cal E}_n\le0$
and $\phi=0$, while the second one describes the case ${\cal E}_n\ge0$ and $\phi=\pi$.

For the first solution at $\chi=\phi=0$ one obtains $t_2=0$. Substituting $t_2=0$ in \eqref{ttt1}, results in
\begin{multline}
t_{1,3}(t_-)=-1\pm\sqrt{1-2{\cal E}_n}=\\ 
=-1\pm\sqrt{(1+f_-^2)^2-2[(g_{n,\delta}+g_\ell)f_--g_\ell f_+(t_-)]^2},
\label{pap24}
\end{multline}
where
\be
{\cal E}_n=-t_--\dfrac12t_-^2+\Bigl[g_\ell f_+-(g_\ell+g_{n,\delta})f_-\Bigr]^2.
\label{pap25}
\ee
It follows from \eqref{pap24} and the condition ${\cal E}_n\!\le0$~that~\mbox{$t_1\!>\!0$.}

The basic dependence $l(t_-)$, that follows from \eqref{sol4} for the first solution at $\chi=\phi=0$, is
\be
l_+(t_-)=\dfrac{2\sqrt{2}}{\sqrt{t_1-t_3}}F\!\left(\left.
\arcsin\sqrt{1-\frac{t_1}{t_-}}\,\right|\,\frac{|t_3|}{t_1+|t_3|}\right),
\label{pap26}
\ee
where the expressions \eqref{pap23}-\eqref{pap25} define $t_{1,3}\equiv f_{1,3}^2$ as a function of
$t_-$.

For the second solution, at $\chi=0$ and $\phi=\pi$, one obtains $t_1=0$. With $t_1=0$ one gets from 
\eqref{ttt1}
\begin{multline}
t_{2,3}(t_-)=-1\pm\sqrt{1-2{\cal E}_n}=\\ 
=-1\pm\sqrt{(1+f_-^2)^2-2[(g_{n,\delta}+g_\ell)f_--g_\ell f_+(t_-)]^2},
\label{pap27}
\end{multline}
It follows from \eqref{pap27} and the condition ${\cal E}_n\ge0$ that both quantities $t_{2,3}\le0$.

The basic dependence $l(t_-)$, that follows from \eqref{sol4} for the second solution at $\chi=0$ and 
$\phi=\pi$, is
\be
l_-(t_-)=\dfrac{2\sqrt{2}}{\sqrt{|t_3|}}F\!\left(\left.
\arcsin\sqrt{\frac{t_-}{t_-+|t_2|}}\,\right|\,\frac{|t_3|-|t_2|}{|t_3|}\right),
\label{pap28}
\ee
where the expressions \eqref{pap23} and \eqref{pap27} define $t_{2,3}\equiv f_{2,3}^2$ as the functions
of $t_-$.

For the first solution, the quantity $t_-(l)$, described by \eqref{pap26}, is a monotonically decreasing
function of $l$, while the second solution \eqref{pap28} monotonically increases with $l$.
The value $t_{-}$ in the limit $l\to0$ is determined for the first solution \eqref{pap26} by 
the equality $t_{-}=t_1$. The latter immediately results in the relation 
\be
f_-=\frac{g_\ell}{g_\ell+g_{n,\delta}}f_+.
\label{rel2}
\ee
For the second solution, one finds from \eqref{pap28} $t_-=0$ in the limit $l\to0$. 

The limiting case $l\to\infty$ corresponds to $t_1=0$ in \eqref{pap26}, and to $t_2=0$ in \eqref{pap28}.
The conditions are equivalent in both cases and take the form 
\be
{\cal E}_n(f_-)=0,
\label{pap29}
\ee
where \eqref{pap23} should be substituted for $f_+$ in \eqref{pap25}. The numerical
studies confirm the above conclusions.

\subsection{Solutions at small distances}
\label{sec: smalll}

The solution of the problem considered can be analytically obtained within the zeroth-order approximation 
in a small parameter $l$. The first argument of the elliptic integral in \eqref{sol4} should vanish in this
case and, therefore, $t_1=t_-$. Taking account of the latter equality in \eqref{ttt1} results in
\begin{align}
t_{2}+t_{3}=&-\left(2+t_-\right),\quad
t_-(t_2+t_3)+t_2t_3=2{\cal E}_n,\nonumber\\
&t_-t_2t_3=2g_{\ell}^2t_-t_+\sin^2\chi.
\label{pap6}
\end{align}

In the limit $l\to0$ the SINIS system is in some aspects similar to symmetric SISIS junctions, and  
the present section generalizes the results of Appendix C in Ref.\,\onlinecite{SBarash2018} to the case, 
when different normalizations in the central and external leads are preferable under different conditions.

There are two solutions to the system of equations \eqref{pap6}. The first solution
is obtained, assuming $t_-\ne0$ and using \eqref{boundcond54} and (5), and results in the relation 
\be
f_-=\dfrac{g_\ell\cos\chi}{g_{n,\delta}+g_\ell}f_+,
\label{rel1}
\ee
which is in agreement with the exact result \eqref{rel2} at $\chi=0$. The condition $g_\ell\cos\chi>0$,
required for  $f_->0$ in \eqref{rel1}, ensures the presence of the proximity effect of the Josephson 
origin.

Substituting \eqref{rel1} for $f_-$ in (5) and in the 
second boundary condition in \eqref{rlambda30a}, taken at $x=l/2+0$, allows one to incorporate the quantities
describing the central electrodes into the effective characteristics of the united interface with boundaries
at $x=\pm (l/2+0)$ in a single symmetric Josephson junction. With the phase incursion over the central lead 
neglected in the limit $l\to0$, the first solution results in 
\be
i=g_\ell^{\text{eff}}f_+^2\sin\phi, \quad g_\ell^{\text{eff}}=\dfrac{g_\ell^2}{2(g_{n,\delta}+g_\ell)}
\label{pap7}
\ee
and in the following equality
\begin{multline}
\dfrac{K}{K_n}\left(\dfrac{df}{dx}\right)_{l/2+0}\!\!\!=
\biggl[\dfrac{g_\delta g_{n,\delta}+g_\ell(g_\delta+g_{n,\delta})}{g_{n,\delta}+g_\ell}+\\ +
2\dfrac{g_\ell^2}{2\left(g_{n,\delta}+g_\ell\right)}\sin^2\dfrac{\phi}{2}\biggr]f_+\, .
\label{calem92}
\end{multline}

The central electrode has mainly been the focus of the study until now, and all the quantities have been 
normalized with respect to the central electrode's characteristics. Since the external superconductor 
electrodes have taken the priority at the moment, it will be convenient to switch over to the normalization 
based on their properties. One introduces 
\begin{align}
&\tilde{f}=\sqrt{\dfrac{a_nb}{|a|b_n}}f,&& \tilde{x}=\sqrt{\dfrac{K_n|a|}{Ka_n}}x,&&
\tilde{l}=\sqrt{\dfrac{K_n|a|}{Ka_n}}l,\nonumber \\
&\tilde{g}_\ell=\sqrt{\dfrac{K_na_n}{K|a|}}g_\ell,&&\tilde{g}_\delta=\sqrt{\dfrac{K_na_n}{K|a|}}g_\delta,&&
\tilde{g}_{n,\delta}=\sqrt{\dfrac{K_na_n}{K|a|}}g_{n,\delta}.
\label{pap8}
\end{align}
and obtains from \eqref{calem92}
\begin{multline}
\left(\dfrac{d\tilde{f}}{d\tilde{x}}\right)_{\tilde{l}/2+0}\!\!\!=
\biggl[\dfrac{\tilde{g}_\delta \tilde{g}_{n,\delta}+\tilde{g}_\ell(\tilde{g}_\delta+
\tilde{g}_{n,\delta})}{\tilde{g}_{n,\delta}+\tilde{g}_\ell}+\\ +
2\dfrac{\tilde{g}_\ell^2}{2\left(\tilde{g}_{n,\delta}+
\tilde{g}_\ell\right)}\sin^2\dfrac{\phi}{2}\biggr]\tilde{f}_+\, .
\label{pap9}
\end{multline}

Equation \eqref{pap9} is of the form of the boundary condition for the order-parameter absolute value 
in a single symmetric Josephson junction with the phase difference $\phi$ across the interface  
\cite{SBarash2012,SBarash2014_3}
\be
\left(\dfrac{d\tilde{f}}{dx}\right)_{l/2+0}=
\Bigl(\tilde{g}_\delta^{\text{eff}}+2\tilde{g}_\ell^{\text{eff}}\sin^2\dfrac{\phi}{2}\Bigr)\tilde{f}_{l/2+0}.
\label{calem41}
\ee

Therefore, the problem of the SINIS junction reduces in the limit $l\to0$ to the behavior of a single 
Josephson junction, described by the first solution. The Josephson current flowing through a single junction 
is known in the GL theory in detail at any coupling constants' values. Here the effective constants of the 
Josephson coupling and the interfacial pair breaking are associated with the characteristics of the SINIS 
junction. The quantity $g_\ell^{\text{eff}}$ is defined in \eqref{pap7}, and $g_\delta^{\text{eff}}$ takes 
the form
\be
g_\delta^{\text{eff}}=\dfrac{g_\delta g_{n,\delta}+g_\ell(g_\delta+g_{n,\delta})}{g_{n,\delta}+g_\ell}.
\label{calem42}
\ee
Changing the normalization will not modify the form of the effective coupling-constants' definitions.

The second solution of the system \eqref{pap6} describes, in the limit $l\to0$, the superconducting
external leads and the normal metal state in the central electrode: $t_1\!=\!t_-\!=\!0$, $i\!=\!0$, 
${\cal E}_n\!=\!g_\ell^2 t_+$, $t_\infty\!=\!\frac{|a|b_n}{a_nb}$, $t_{2,3}=-1\pm\sqrt{1-2g_\ell^2 t_+}$,
\be
t_+=\!\biggl(\!\sqrt{\frac{|a|b_n}{a_nb}+\frac{K_nb_n}{2Kb}(g_\delta+g_\ell)^2}-
\sqrt{\frac{K_nb_n}{2Kb}}(g_\delta+g_\ell)\!\biggr)^2\!. 
\label{pap10}
\ee
The solution does not contain any phase dependence as no 
superconductivity is present in the central electrode.

\subsection{Solutions with small $f_-$ for tunnel interfaces}
\label{sec: tunnel}

The analytical solutions to the GL equation can be found assuming $f_-\sim g_\ell f_+\ll1$ and 
$g_{n,\delta}\alt g_\ell$. The former condition makes possible to disregard the cubic term in the first GL 
equation in \eqref{rlambda11}, as compared to the linear one:
\be
\dfrac{d^2f}{dx^2}-\dfrac{i^2}{f^3}-f=0,\quad |x|<l/2.
\label{pap11}
\ee
The solution to this equation is
\be
f(x)=\sqrt{t_1\cosh^2x-t_2\sinh^2x},
\label{pap12}
\ee
where the parameters $t_1\ge0$ and $t_2\le0$ satisfy
\be
t_1t_2=-i^2, \quad t_1+t_2=-{\cal E}_n.
\label{pap13}
\ee
The absolute value of the order parameter inside the central lead has the maximums $f_-$ at its side faces
$x=\pm l/2$ and the minimum $f_1$ at its center $x=0$.

Assuming $g_\ell f_+$ and $f_-$ to be the quantities of the same order of smallness, one 
can satisfy the relation $f_-\gg 2g_\ell(g_\ell+g_{n,\delta})f_+$ and reduce
the expression \eqref{boundcond54} for ${\cal E}_n$ to the following simplified form
\be
{\cal E}_n=-t_-+g_\ell^2 t_+.
\label{pap14}
\ee

Using \eqref{pap12}-\eqref{pap14} and (5), one gets the system of equations
\begin{align}
&t_1t_2=-g^2_{\ell}t_-t_+\sin^2\chi_r, \nonumber\\
&t_1+t_2=t_--g^2_{\ell}t_+, \nonumber\\
& t_-=t_1\cosh^2\frac{l}{2}-t_2\sinh^2\frac{l}{2},
\label{pap15}
\end{align}
where the quantities $t_-$, $t_{1,2}$ are on the order of $g^2_{\ell}t_+$, with the higher 
order terms disregarded.

There are two solutions to \eqref{pap15}:
\begin{widetext}
\begin{align}
&f_{-,\pm}=g_{\ell}f_+\coth l\biggl[\cos\chi \pm \sqrt{\cos^2\chi-\tanh^2 l}\,\biggr],
\label{lsmall132}\\
&t_{1,\pm}=\frac12 g^2_{\ell} t_+
\biggl\{-2+\biggl(1+\coth^2\frac{l}2\biggr)\cos\chi\biggl[\cos\chi\pm \sqrt{\cos^2\chi-\tanh^2 l}\,
\biggr]\biggr\},
\label{lsmall133}\\
&t_{2,\pm}=\frac12 g^2_{\ell} t_+
\biggl\{-2+\biggl(1+\tanh^2\frac{l}2\biggr)\cos\chi\biggl[\cos\chi\pm \sqrt{\cos^2\chi-\tanh^2 l}\,
\biggr]\biggr\},
\label{lsmall134}
\end{align}
\end{widetext}
taking place under the condition 
\be
|\chi|\le\chi_{\text{max}}(l)=\arccos\tanh l.
\label{chimax}
\ee

Within the same approximation one obtains that the last term on the right hand side in \eqref{calem10p}
can be disregarded and $f_\infty$ coincides with its currentless value.
The results can be represented as
\be
\tilde{f}_\infty=1, \quad \tilde{f}_+=1-\dfrac1{\sqrt{2}}(\tilde{g}_\delta+\tilde{g}_\ell),
\label{pap16}
\ee
where the normalization defined in \eqref{pap8} is used.
Since $t_+$ should be taken in \eqref{lsmall132}-\eqref{lsmall134} within the zeroth approximation in 
$g_\ell$, one has $t_+^{(0)}=\frac{b_n|a|}{ba_n}$.

The quantity $f_{-,+}(\chi,l)$, i.e., the first solution for $f_-$ in \eqref{lsmall132} taken at 
the side face, has its maximum at $\chi=0$ and minimum at $\chi=\pm\chi_{\text{max}}(l)$. At the same 
time, the second solution $f_{-,-}(\chi,l)$ for $f_-$ has its maximum at $\chi=\chi_{\text{max}}(l)$ 
and minimum at $\chi=0$:
\begin{align}
&f_{-,+}(0,l)=g_\ell\coth\frac{l}2\left(\frac{b_n|a|}{ba_n}\right)^{1/2}, \label{pap17}\\
&f_{-,-}(0,l)=g_\ell\tanh\frac{l}2\left(\frac{b_n|a|}{ba_n}\right)^{1/2}, \label{pap18}\\
&f_{-,\pm}(\chi_{\text{max}}(l),l)=g_\ell\left(\frac{b_n|a|}{ba_n}\right)^{1/2}.
\label{pap19}
\end{align}
It is worth noting that the quantity $f_{-,\pm}(\chi_{\text{max}}(l),l)$ does not depend on $l$
within the approximation used. The limiting behavior $f_{-,-}(0,l)\to0$ at $l\to0$ agrees with
the exact result \eqref{pap28} and the second solution in Sec.\,\ref{sec: smalll}.

Substituting the maximal order parameter value \eqref{pap17} in the presumed condition $f_-\ll1$,
one finds that the first solution in  \eqref{pap12}, \eqref{lsmall132}-\eqref{lsmall134}, taken at 
$\cos\chi\sim1$, is justified when the length of the
central lead is not too small: $l\gg 2g_\ell\left(\frac{b_n|a|}{ba_n}\right)^{1/2}$. For the parameter set 
used in the main text in plotting the figures, one gets $l\gg 0.02$. At sufficiently  small $l$ and 
$\cos\chi\sim1$  the first solution has been considered in the preceding section. At the same time, 
the results obtained in this section can be well applied at any $l$ to the second solution, as well
as to the first one in a vicinity of $\chi=\chi_{\text{max}}(l)$. 

For the absolute order parameter value $f_{1}$, at the center of the normal metal lead $x=0$, one finds from
\eqref{lsmall133}:
\begin{align}
&f_{1,+}(0,l)=\dfrac{g_\ell}{\sinh\frac{l}2}\left(\frac{b_n|a|}{ba_n}\right)^{1/2}, \label{pap20}\\
&f_{1,-}(0,l)=0, \label{pap21}\\
&f_{1,\pm}(\chi_{\text{max}}(l),l)=\dfrac{g_\ell}{\sqrt{\cosh l}}\left(\frac{b_n|a|}{ba_n}\right)^{1/2}.
\label{pap22}
\end{align}

The most interesting point regarding $f_{1}$ is the vanishing second solution at $\chi=0$. In other words,
in agreement with earlier results \cite{SFink1976,SGiazotto2017}, the order parameter takes zero value
at the center of the normal metal lead that makes possible phase-slip processes at $\phi=\pi$, and at
arbitrary $l$. For the values $\chi=0$ and $\phi=\pi$ the phase incursion is $\varphi=\pi$. As 
the supercurrent vanishes at $\chi=0$ and the phase gradient satisfies the relation \eqref{pap3},
the phase incursion
can differ from zero in the limit $i\to0$ only if the order parameter also vanishes in this limit somewhere
inside the central electrode. This has to be the point $x=0$, since it is the minimum of $f(x)\ge0$.

The expression for the phase incursion $\varphi$, that follows from \eqref{pap3} with the solutions 
\eqref{pap12}, \eqref{lsmall132} and \eqref{lsmall133}, is
\begin{align}
&\varphi_\pm(\chi,l)=\sgn(\sin\chi)\arccos\biggl[\cosh l\biggl(\sin^2\chi\pm \nonumber \\
&\qquad \qquad \qquad \qquad \pm\cos\chi\sqrt{\cos^2\chi-\tanh^2 l}\biggr)\biggr].
\label{lsmall158}
\end{align}
In particular, one gets $\varphi_+(0,l)=0$ for the first solution, $\varphi_-(0,l)=\pi$
for the second one and $\varphi_\pm(\chi_{\text{max}}(l),l)=\arccos(1/\cosh l)$, where 
$\chi_{\text{max}}(l)$ is defined in \eqref{chimax}. Two possible values $0$ and $\pi$ 
of $\varphi$ in currentless states of the double junctions have been earlier identified 
in Refs.\,\onlinecite{SVolkovA1971} and \onlinecite{SFink1976}.

As the external phase difference is determined by the relation $\phi(\chi,l)=2\chi+\varphi_\pm(\chi,l)$,
one obtains
\be
\sin\phi_\pm(\chi,l)=
\Biggl[\cos\chi\pm \sqrt{\cos^2\chi-\tanh^2 l}\Biggr]\cosh l\sin\chi.
\label{lsmall168}
\ee
For $\phi_*(l)\equiv\phi(\chi_{\text{max}}(l),l)$ one gets from \eqref{lsmall168}:
\be
\phi_*(l)=\dfrac{\pi}{2}+\arcsin(1/\cosh l).
\label{pap30}
\ee

The sinusoidal current-phase relation follows from (5), \eqref{lsmall132} and \eqref{lsmall168}:
\be
i=\dfrac{g_\ell^2}{\sinh l}t_+\sin\phi.
\label{pap31}
\ee 
Using \eqref{pap16} and the normalization \eqref{pap8}, associated with the external superconducting leads, 
the supercurrent can be approximately written as
\be
\tilde{i}=\tilde{i}_c\sin\phi, \quad \tilde{i}_c=\dfrac{g_\ell\tilde{g}_\ell}{\sinh l}.
\label{pap32}
\ee
The condition $\tilde{i}_c\ll1$, which should hold for tunnel junctions, results in 
$l\gg g_\ell\tilde{g}_\ell$.

A comparison with the numerical results show that the approximate description of the phase relations, which 
is based on \eqref{lsmall158}, \eqref{pap30} has a noticeably better accuracy than the description of the 
order parameter with \eqref{lsmall132}-\eqref{lsmall134}, obtained within the same approximation. A possible 
reason for that can be associated with different character of the higher-order corrections to those results.
It is also worth noting that, within the approximation used, the coupling constants have canceled out in 
\eqref{lsmall158}-\eqref{pap30}, as opposed to \eqref{lsmall132}-\eqref{lsmall134}.

\onecolumngrid
\vspace{-1ex}


\end{document}